\documentclass{PoS}

\title{Towards the understanding of jet substructures and cross sections in heavy ion collisions using soft collinear effective theory}

\ShortTitle{Jet substructure and cross section in heavy ion collisions using SCET}

\author{\speaker{Yang-Ting Chien}\\%\thanks{}\\
        Theoretical Division, T-2, Los Alamos National Laboratory, Los Alamos, NM 87545\\
        E-mail: \email{ytchien@lanl.gov}}

\abstract{The jet quenching phenomenon in heavy ion collisions provides a strong evidence of the modification of parton shower in the quark-gluon plasma (QGP). Jet substructure observables can probe various aspects of the jet formation mechanism. They contain useful information about the QGP and allow us to study the medium properties in great details. Here we present theoretical calculations of jet shapes and cross sections in proton-proton and lead-lead collisions at the LHC using soft-collinear effective theory, with Glauber gluon interactions in the medium. We find that resumming large logarithms in the jet substructure calculation is necessary for precise theoretical predictions. The resummation is performed using renormalization group evolution between characteristic jet scales. We also find that the medium induces power corrections to jet shapes. In the end we present the comparison between our calculations with the recent measurements at the LHC with very good agreement. Our calculations help initiate precise jet modification studies in heavy ion collisions.}

\FullConference{38th International Conference on High Energy Physics\\
		3-10 August 2016\\
		Chicago, USA}

\usepackage{url,overcite}
\usepackage{multicol,psfrag,cite,cancel}
\usepackage[svgnames]{xcolor}

\newcommand{\be}{\begin{equation}}
\newcommand{\ee}{\end{equation}}
\newcommand{\bea}{\begin{eqnarray}}
\newcommand{\eea}{\end{eqnarray}}

\begin{document}

\section{Introduction}

The strong suppression of hadron and jet production cross sections in heavy ion collisions at the Relativistic Heavy Ion Collider (RHIC) and the Large Hadron Collider (LHC) has long been observed which establishes the phenomenon of jet quenching and the creation of the quark-gluon plasma (QGP). Although the suppression of cross sections could be described quite well by several models exploiting the parton energy loss picture, it has been clear that more differential and correlated measurements are needed in order to distinguish various jet formation mechanisms.

Jet substructure observables can resolve jets at different energy scales. They are more sensitive to the final-state, jet-medium interaction which allows us to separate the initial-state effects that can directly affect the production cross sections. The interference between jets and the medium makes jet physics in heavy ion collisions an even more complicated multi-scale problem. To this end, effective field theory techniques are extremely useful in separating physics at multiple characteristic scales. In this talk I will present the precise calculation of the jet shape \cite{Ellis:1992qq} in both proton-proton \cite{Chien:2014nsa} and lead-lead \cite{Chien:2015hda} collisions using soft-collinear effective theory (SCET) \cite{Bauer:2000ew, Bauer:2000yr, Bauer:2001ct, Bauer:2001yt, Bauer:2002nz, Beneke:2002ph}, with Glauber gluon interactions in the medium \cite{Idilbi:2008vm, Ovanesyan:2011xy}. The same framework has also been successfully applied in the calculation of jet fragmentation functions \cite{Chien:2015ctp}.

%I will introduce the jet shape and discuss its factorization theorem in SCET. The resummation of the jet shape is performed at next-to-leading logarithmic (NLL) accuracy using renormalization-group techniques. In heavy ion collisions the medium modification to the jet shape as well as the jet energy loss are calculated using the medium-induced splitting functions \cite{Ovanesyan:2011kn, Fickinger:2013xwa}. In the end I will compare the theoretical calculation with the data from ALICE \cite{Abelev:2013kqa}, ATLAS \cite{Aad:2012vca,Aad:2014bxa} and CMS \cite{Chatrchyan:2013kwa,CMS:prelim} and make predictions for the upcoming Run 2 measurements.

%\section{The jet shape}
%\label{sec:obs}
\section{The factorization theorem of the jet shape in SCET}
\label{sec:fac}

%\begin{figure}[top]
%\center
%\psfrag{x}{\footnotesize $r$}
%\psfrag{z}{\footnotesize $\rho(r)$}
%    \includegraphics[height=4cm, trim = 10mm -15mm 0mm 0mm]{../../Talks/Figures/jetshape.eps}~~~~~~~~~~~~~~~
%    \includegraphics[height=4cm, trim = 0mm 0mm 0mm 0mm]{../../Talks/Figures/Djetshape1}
%\caption{Left: The jet shape probes the transverse energy distribution within a reconstructed jet as a function of the subcone size $r$. Right: The resummed (solid) and fixed-order (dashed) differential jet shapes of quark-initiated (blue) and gluon-initiated (red) jets. }
%\end{figure}
The jet shape \cite{Ellis:1992qq} probes the transverse energy distribution inside a reconstructed jet with radius $R$. It is defined as the fraction of the transverse energy $E_T$ of the jet within a subcone of size $r$,
\be
    \Psi_J(r)=\frac{\sum_{r_{i}<r} E^i_T}{\sum_{r_{i}<R} E^i_T}\;.
\ee
This quantity is averaged over all jets and the derivative $\rho(r)$ describes how the transverse energy is differentially distributed in $r$.
\be
    \Psi(r)=\frac{1}{N_J}\sum_{J=1}^{N_J} \Psi_J(r),~~~~~~~~\rho(r)=\frac{d}{dr}\Psi(r)\;.
\ee
%In heavy ion collisions, the modification of the jet shape is conventionally quantified by taking the ratio of the jet shapes in nucleus-nucleus and proton-proton collisions $\rho^{AA}(r)/\rho^{pp}(r)$.
The infrared structure of QCD induces Sudakov logarithms of the form $\alpha_s^n\log^m r/R~(m\leq 2n)$ in the perturbative calculation of the jet shape. The fixed order calculation breaks down at small $r$ and the large logarithms need to be resummed. %I will demonstrate how the all-order resummation of the jet shape is performed using SCET.

%\begin{figure}[top]
%\center
%\includegraphics[width=0.26\linewidth]{../../Talks/Figures/SCET}~~~~~~~~~~~~~~
%\includegraphics[width=0.28\linewidth]{../../Talks/Figures/factorization}
%\caption{Left: The factorization of the {\color{blue}hard}, {\color{green!50!black}collinear} and {\color{red}soft} sectors in SCET. Right: The schematic illustration of the jet production and the measurement of the jet shape.}
%\label{fig:SCET}
%\end{figure}

Effective field theory techniques are useful whenever there is clear scale separation. SCET separates the physical degrees of freedom in QCD by a systematic expansion in power counting. In events with the production of energetic and collimated jets, the power counting parameter is small and the leading-power contribution in SCET is a very good approximation of the full QCD result. We first match SCET to QCD at the hard scale by integrating out the contributions from the hard modes. We then integrate out the off-shell modes which gives collinear Wilson lines describing the collinear radiation. The soft sector is described by soft Wilson lines along the jet directions. SCET factorizes a complicated, multi-scale problem into multiple simpler, single-scale problem \footnote{The soft function of exclusive observables may still be multi-scaled and need to be refactorized \cite{Ellis:2010rwa,Becher:2015hka,Chien:2015cka}},
which allows us to calculate the contribution to the physical cross section from each sector separately.

The factorization theorem for the differential cross section of the production of $N$ jets with transverse momenta $p_{T_i}$ and rapidity $y_i$, the energy $E_r$ inside the cone of size $r$ in one jet, and an energy cutoff $\Lambda$ outside all the jets is the following,
\be
    \frac{d\sigma}{dp_{T_i}dy_i dE_r}
    =H(p_{T_i},y_i,\mu)J_1^{\omega_1}(E_r,\mu) J_2^{\omega_2}(\mu)\dots J_N^{\omega_N}(\mu)S_{1,2,\dots N}(\Lambda,\mu).
\ee
$H(p_{T_i},y_i,\mu)$ is the hard function describing the hard scattering process at high energy. $J^{\omega}(E_r,\mu)$ is the jet function which is the probability of having the energy $E_r$ inside a subcone of size $r$ in the jet with energy $\omega=2E_J$. All the other jet functions $J^{\omega}(\mu)$ are unmeasured jet functions \cite{Ellis:2010rwa} without measuring the substructure of the jet. $S_{1,2,\dots N}(\Lambda,\mu)$ is the soft function and it describes how soft radiation is constrained in measurements. The factorization theorem simplifies dramatically and has a product form. The averaged energy inside the cone of size $r$ in jet 1 is the following,
\be
    \langle E_r\rangle_{\omega_1}
    =\frac{\int dE_r E_r\frac{d\sigma}{dp_{T_i}dy_idE_r}}{\frac{d\sigma}{dp_{T_i}dy_i}}
    =\frac{H({p_T}_i,y_i,\mu)J^{\omega_1}_{{E,r}_1}(\mu)J^{\omega_2}_2(\mu)\dots S_{1,2,\dots}(\Lambda,\mu)}{H({p_T}_i,y_i,\mu)J^{\omega_1}_1(\mu)J^{\omega_2}_2(\mu)\dots S_{1,2,\dots}(\Lambda,\mu)}
            =\frac{ J^{\omega_1}_{{E,r}_1}(\mu)}{J^{\omega_1}_1(\mu)}\;,
\ee
and $J_{E,r}^{\omega}(\mu)=\int dE_r E_r~J^\omega(E_r,\mu)$ is referred to as the jet energy function. Note the huge cancelation between the hard, unmeasured jet and soft functions. This implies that the jet shape is insensitive to the underlying hard scattering process as well as the other part of the event. The integral jet shape is weighted with the jet production cross sections with proper phase space cuts on $p_T$ and $y$,
\be
    \Psi(r)=\frac{1}{\sigma_{\rm total}}\sum_{i=q,g}\int_{PS} dp_Tdy \frac{d\sigma^{i}}{dp_Tdy}\Psi^i_\omega(r)\;,~{\rm where}~
    \Psi_\omega(r)
    =\frac{J_{E,r}(\mu)/J(\mu)}{J_{E,R}(\mu)/J(\mu)}
    =\frac{J_{E,r}(\mu)}{J_{E,R}(\mu)}\;.
\ee
The jet shape is a collinear observable and is relatively insensitive to the soft radiation.

\section{Scale hierarchy and renormalization-group evolution}
\label{sec:RG}

%\begin{figure}[top]
%\center
%\includegraphics[height=3cm]{../../Talks/Figures/jetshape}~~~~~~
%\includegraphics[height=3cm]{../../Talks/Figures/scale}
%\caption{The renormalization-group evolution of the jet energy function between $\mu_{j_r}$ and $\mu_{j_R}$ resums $\log \mu_{j_r}/\mu_{j_R}=\log r/R$.}
%\label{fig:RG}
%\end{figure}

The renormalization-group evolution of the jet energy function allows us to resum the jet shape. It satisfies the following RG equation ,
\be
    \frac{d J^{q}_{E,r}(r,R,\mu)}{d\ln\mu}
    =\left[-C_F~\Gamma_{\rm cusp}(\alpha_s)\ln\frac{\omega^2\tan^2\frac{R}{2}}{\mu^2}-2\gamma_{J^q}(\alpha_s)\right]J^{q}_{E,r}(r,R,\mu)\;,
\ee
for quark jets (for gluon jets with the color factor $C_A$), and $\Gamma_{\rm cusp}$ is the cusp anomalous dimension. Note that the integral jet shape $\Psi_\omega$ is renormalization group invariant,
\be
    \Psi_\omega=\frac{J_{E,r}(\mu)}{J_{E,R}(\mu)}=\frac{J_{E,r}(\mu_{j_r})}{J_{E,R}(\mu_{j_R})}U_J(\mu_{j_r},\mu_{j_R})\;.
\ee
The scale $\mu_{j_r}=\omega\tan\frac{r}{2}\approx E_J \times r$ can be identified \cite{Chien:2014nsa} which eliminates the large logarithms in the fixed-order calculation of the jet energy function $J_{E,r}(\mu_{j_r})$. The hierarchy between $r$ and $R$ induces two hierarchical jet scales $\mu_{j_r}$ and $\mu_{j_R}$, and $U_J(\mu_{j_r},\mu_{j_R})$ is the RG evolution kernel which resums the logarithms of the ratio between $\mu_{j_r}$ and $\mu_{j_R}$ $(\log\mu_{j_r}/\mu_{j_R}=\log r/R)$.

%\section{Baseline jet shape calculations}
%\label{sec:pp}

\begin{figure}[top]
\center
\psfrag{x}{$r$}
\psfrag{z}{$\rho(r)$}
\includegraphics[width=0.4\linewidth]{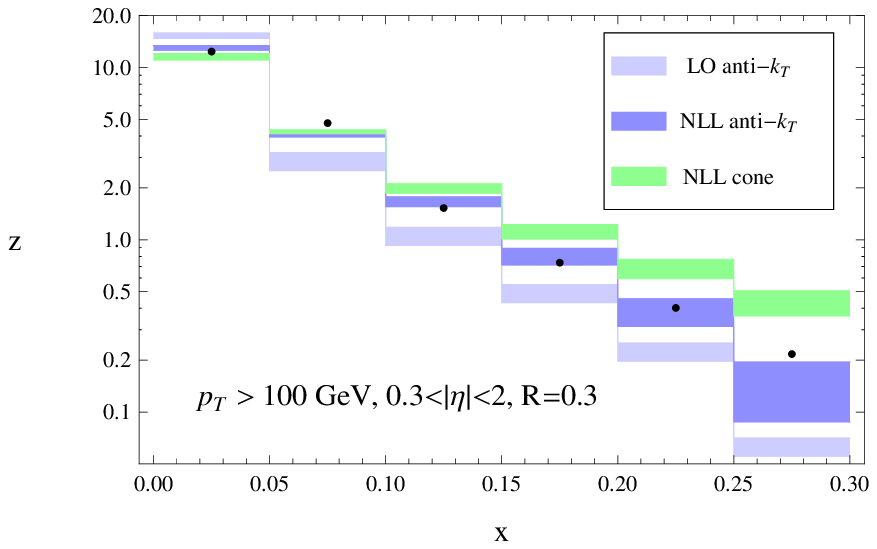}~~~~
\includegraphics[width=0.4\linewidth]{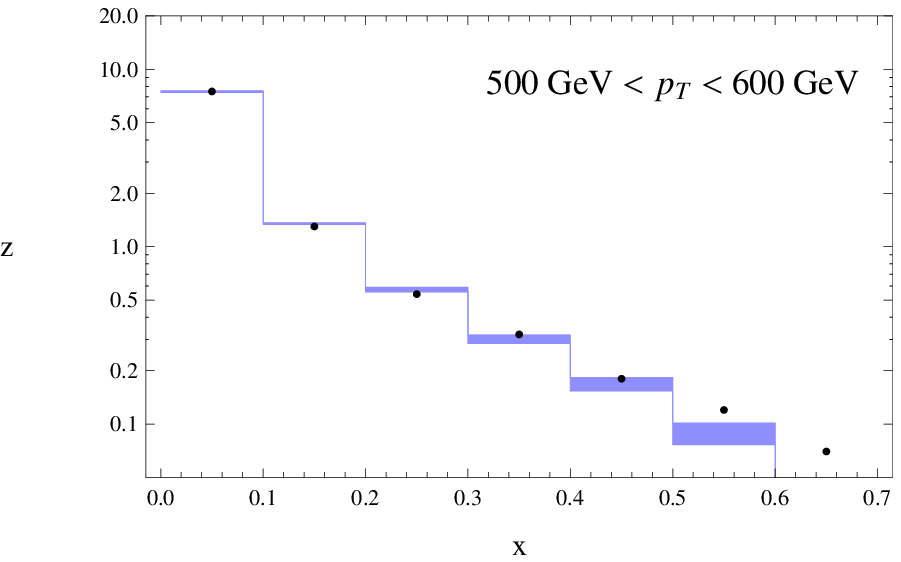}
\caption{Left: The differential jet shape for anti-$\rm k_T$ jets with $R=0.3$, $p_T>100$ GeV and $0.3<|y|<2$ in proton-proton collisions at the 2.76 TeV LHC. The black dots are the data from CMS. The shaded blue boxes are the LO (light) and NLL (dark) calculations, while the shaded green boxes are the NLL calculation for cone jets. Right: The differential jet shape for anti-$\rm k_T$ jets with $R=0.7$ at the 7 TeV LHC.}
\label{fig:resum}
\end{figure}

We compare our calculations with the CMS measurements at the 2.76 TeV \cite{Chatrchyan:2013kwa} and 7 TeV LHC \cite{Chatrchyan:2012mec} (FIG.\ref{fig:resum}). The resummed NLL calculation agrees much better with data than the fixed-order result. Bands are theoretical uncertainties estimated by varying the jet scales $\mu_{j_r}$ and $\mu_{j_R}$ in the resummed expressions. The shape difference for jets reconstructed using the anti-$k_T$ and the cone algorithms is significant. In the region $r\approx R$, higher fixed-order calculations and power corrections become more prominent.

\section{Multiple scattering and $\rm SCET$ with Glauber gluons}

Coherent multiple scattering and the induced bremsstrahlung are the qualitatively new features of the in-medium parton shower evolution. The Debye screening scale $\mu$ sets the range of interaction, and the parton mean free path $\lambda$ determines the significance of multiple scattering in the medium. The radiation formation time $\tau$ sets the scale where the jet and the medium can be resolved. Parton splitting and induced bremsstrahlung interfere in the jet formation, and the interplay among these characteristic scales can result in different interference patterns.

We identify the Glauber gluon as the relevant mode describing the momentum transfer transverse to the jet direction between the jet and the medium, and the extended effective theory is dubbed $\rm SCET_G$ \cite{Idilbi:2008vm, Ovanesyan:2011xy}. The Glauber gluons are generated from the color charges in the medium. Given a medium model, we can consistently couple the medium to jets using $\rm SCET_G$. From thermal field theory and lattice QCD calculations, an ensemble of quasi particles with Debye screened potential and thermal masses is a reasonably valid parameterization of the medium properties. We adopt this medium model with the Bjorken-expanded hydrodynamic evolution \cite{Bjorken:1982tu}.

%\begin{figure}[top]
%\center
%\includegraphics[height=2.5cm, trim = 0mm 0mm 0mm 8mm]{../../Talks/Figures/shower.eps}~~~
%\includegraphics[height=4cm]{../../Talks/Figures/SCETG.eps}~~~
%\includegraphics[height=2.5cm, trim = 0mm 0mm 0mm 2mm]{../../Talks/Figures/glauber.eps}
%\caption{Left: Illustration of the emergent scales in the medium jet formation. Middle: The factorization of the {\color{blue}hard}, 5{\color{green!50!black}collinear}, {\color{red}soft} and {\color{green!50!black}Glauber} sectors in $\rm SCET_G$. Right: Parton splitting kinematics at leading order.}
%\label{fig:medium}
%\end{figure}

%\section{Medium-induced splitting function and Landau-Pomeranchuk-Migdal effect}

The large angle bremsstrahlung takes away energy, resulting in the jet energy loss and the modification of the jet shape. The key ingredients which enter the calculations of the medium modification of the jet shape are the medium-induced splitting functions \cite{Ovanesyan:2011kn,Fickinger:2013xwa}. The jet shape can be calculated using the collinear parton splitting functions. At leading order,
\be
    J^{i}_{E,r}(\mu)=\sum_{j,k}\int_{PS} dxdk_\perp\Big[\frac{dN^{vac}_{i\rightarrow jk}}{dxd^2k_\perp}+\frac{dN^{med}_{i\rightarrow jk}}{dxd^2k_\perp}\Big] E_r(x,k_\perp)\;.
\ee
The jet shape in heavy ion collisions gets modified through the modification of jet energy functions,
\be
    \Psi(r)
    =\frac{J^{vac}_{E,r}+J^{med}_{E,r}}{J^{vac}_{E,R}+J^{med}_{E,R}}
    =\frac{\Psi^{vac}(r)J^{vac}_{E,R}+J^{med}_{E,r}}{J^{vac}_{E,R}+J^{med}_{E,R}}\;.
\ee
Here $J_{E,r}^{med}(r)$ contributes as a power correction without large logarithms. There is no extra soft-collinear divergence and the RG evolution of the jet energy function is the same as in vacuum. The jet shape is then averaged with the jet cross section which is significantly suppressed due to the jet energy loss in heavy ion collisions \cite{Chien:2015hda}. Note that gluon-initiated jets are quenched more than quark-initiated jets therefore the quark jet fraction is increased.

\begin{figure}[top]
\center
\psfrag{x}{$r$}
\psfrag{w}{\small $\frac{\psi(r)^{\rm Pb}}{\psi(r)^{\rm P}}$}
    \includegraphics[height=4cm]{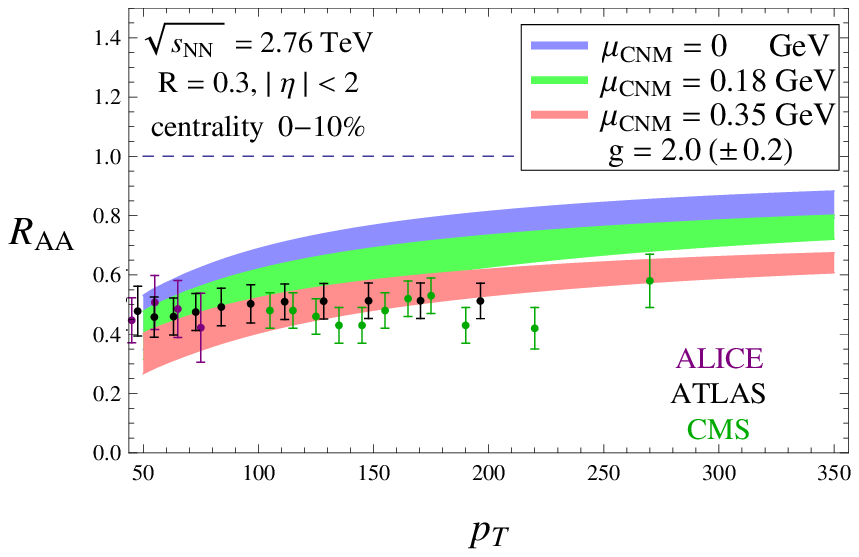}~~~~~
    \includegraphics[height=4.2cm]{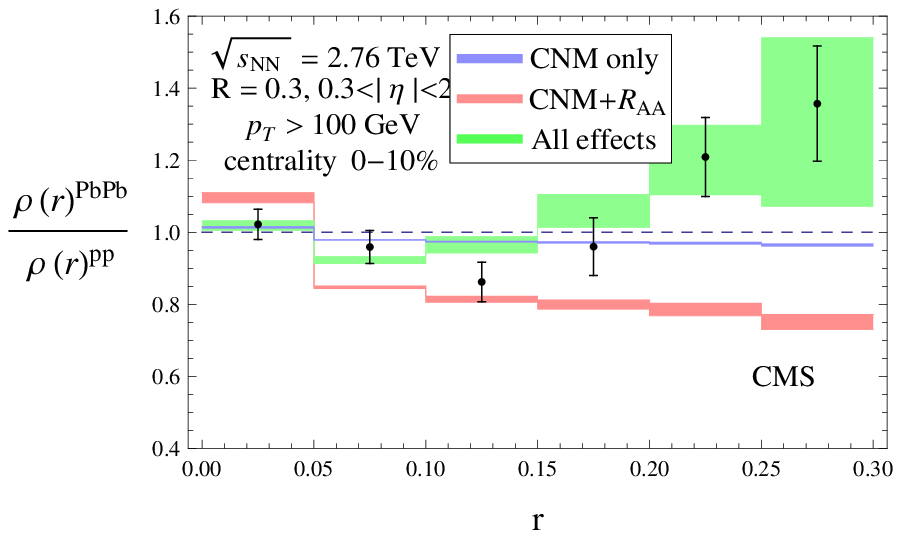}
\caption{Left: Nuclear modification factor $R_{AA}$ for anti-$k_T$ jets with $R=0.3$, $|\eta|<2$ as a function of $p_T$ in central lead-lead collisions at $\sqrt{s_{\rm NN}}=2.76$ TeV, with different cold nuclear matter effects implemented. Right: The ratio of the jet shapes in pp and PbPb collisions, with the CNM effect (blue), plus the cross section suppression (red), plus the jet-by-jet shape modification (green).}
\label{fig:result1}
\end{figure}

We present the theoretical calculations of the nuclear modification factor $R_{AA}$ for jet cross sections as well as the ratio of the differential jet shapes in lead-lead versus proton-proton collisions at $\sqrt{s_{\rm NN}}=2.76$ TeV (FIG.\ref{fig:result1}). The left plot shows the results with different CNM effects implemented in the calculations, and we compare with the measurements from ALICE \cite{Abelev:2013kqa}, ATLAS \cite{Aad:2012vca} and CMS \cite{CMS:prelim}. The theoretical uncertainty is estimated by varying the jet-medium coupling $g$. We see that the jet cross section is quite sensitive to the CNM effect. The right plot shows the jet shape calculation and studies its sensitivity to the CNM effect, the quark/gluon jet fraction and the jet-by-jet shape modification. The shaded boxes represent the scale uncertainty. We see that the non-trivial jet shape modification pattern observed at CMS \cite{Chatrchyan:2013kwa} is due to both the increase of the quark jet fraction, which tends to make the jet shape narrower, as well as the broadening of the jet-by-jet shape, and it is insensitive to the CNM effect. We also examine the dependence of $R_{AA}$ on centrality, the jet rapidity and the jet radius \cite{Aad:2012vca,Aad:2014bxa} and make predictions for the jet shape and cross section of inclusive and photon-tagged jets at the LHC Run 2 \cite{Chien:2015hda}.

\section{Conclusions}

The jet shape and cross section in proton and heavy ion collisions are calculated using the SCET extended with Glauber gluon interactions. The jet shape is resummed at NLL accuracy using renormalization-group techniques, and the baseline calculation is established. The medium modification to the jet shape is calculated using the medium-induced splitting functions. We find good agreement between our calculations. The physics understanding is that the non-trivial jet shape modification pattern is due to the combination of the cross section suppression and the jet-by-jet broadening. The precise jet substructure studies in heavy ion collisions have been initiated and we are entering the golden age.

%\begin{figure}[top]
%    \psfrag{x}{$r$}
%    \psfrag{w}{\small $\frac{\psi(r)^{\rm Pb}}{\psi(r)^{\rm P}}$}
%        \includegraphics[height=3.0cm]{../../Talks/Figures/RAA_ATLAS_cent-fin}
%        \includegraphics[height=3.1cm]{../../Talks/Figures/RAA_ATLAS_vsY-fin}
%        \includegraphics[height=3.1cm]{../../Talks/Figures/RCP_RCP02_ATLAS-fin}
%\caption{Examination of the centrality (left), the jet rapidity (middle) and the jet radius (right) dependence in the jet cross section suppression. The theory calculations agree with the data very well.}
%\label{fig:result2}
%\end{figure}

\section{Acknowledgments}

This work is supported by the US Department of Energy, Office of Science.

%\begin{figure}[t]
%\psfrag{x}{$r$}
%    \psfrag{w}{\small $\frac{\psi(r)^{\rm Pb}}{\psi(r)^{\rm P}}$}
%        \includegraphics[height=3.1cm]{../../Talks/Figures/RAA_INCLvsPHOT_R04-fin}
%        \includegraphics[height=3.1cm]{../../Talks/Figures/Shape_INCLvsPHOT_R03-fin}
%        \includegraphics[height=3.1cm]{../../Talks/Figures/Rshape_INCLandPHOT_010_R03-fin}
%\caption{Predictions for the jet cross section (left) and the jet shape (middle and right) at the $\sqrt{s_{\rm NN}}=5$ TeV LHC for inclusive and photon-tagged jets. Due to the higher quark jet fraction, for photon-tagged jets the cross section suppression is less and and the jet broadening is more manifest.}
%\label{fig:result3}
%\end{figure}

\end{document}